%% file: DFC-OSID-arXiv.tex
\documentclass[aps,showpacs,amssymb,amsfonts,superscriptaddress,pra,onecolumn,nofootinbib,10pt]{revtex4-1}
\usepackage{bm,bbm}
\usepackage{times}
\usepackage{graphicx,amsthm,epstopdf,amsmath}
\usepackage{subfigure,color}
\usepackage[draft]{fixme}
\usepackage[super]{nth}
\usepackage[top=1.25in, bottom=1.25in, left=0.70in, right=0.70in]{geometry}
\input{komendy_osid.tex}

\newtheorem{thm}{Theorem}

\newtheorem{lem}[thm]{Lemma}
\newtheorem{rem}[thm]{Remark}
\newtheorem{defi}[thm]{Definition}

\newtheorem{cor}[thm]{Corollary}

\begin{document}

\title{Decoherence-Free Communication over Multiaccess Quantum Channels}
\author{Maciej Demianowicz}
\affiliation{Atomic Physics Division, Department of Atomic Physics and Luminescence, 
Faculty of Applied Physics and Mathematics,
Gda\'nsk University of Technology \\ ul. Narutowicza 11/12, PL80-233 Gda\'nsk, 
Poland  \\[1ex] and \\[1ex] 
National Quantum Information Center in Gda\'nsk, ul. W\l. Andersa 27,
PL81---824 Sopot, Poland }

\begin{abstract}In this paper we consider decoherence-free communication over
multiple access and $k$-user quantum channels. First, we
concentrate on a hermitian unitary noise model $U$ for a
 two-access bi-unitary channel and show that in this case a
decoherence-free code exists if the space of Schmidt matrices of
an eigensubspace of $U$ exhibits certain properties of
decomposability. Then, we show that our technique is also
applicable for generic random unitary  two-access channels.
Finally, we consider the applicability of the result to the case of a
larger number of senders and general Kraus operators.
\end{abstract}

\maketitle

\section{Introduction}

Quantum information transmission 
\cite{bennett-huge,lloyd,barnum-nielsen-schumacher,BKN,devetak} 
through a nontrivial quantum channel inevitably involves the occurrence of errors. 
These errors, handled in an incompetent way,  may completely shadow an 
intended quantum message. In this context, methods of combating such errors naturally emerge as one of the main topics of the theory of quantum channels.
Luckily, several useful techniques have been developed to overcome destructive influence of
 coupling to the environment (see, e.g., \cite{byrde-et-al,kribs-et-al}). Among them,
 quantum error correction codes (QECC) are most widely recognized (see \cite{gaitan-ksiazka} and references therein).
Methods of constructing QECC for quantum communication have been previously reported in the
 literature  \cite{steane,calderbank-shor,gottesman,choi-et-al}. 
However, these proposals concerned
bipartite communication (see however \cite{wilde}). No general approach has been developed
 to treat the case of the larger number of users of a quantum network.

The present work is a step in an effort towards one such general approach. We concentrate on random
unitary multiaccess channels and show how to construct decoherence-free codes for such channels.
  Decoherence-free codes are the carriers of quantum information on which it is completely safe from the influence of an environment,
  that is a quantum message goes undisturbed, in the sense of a solely unitary operation, through a channel.
  Since the recognition of their significance, they have drawn much attention (see, \np, \cite{dfs} and
  references therein) and the concept found its experimental realizations \cite{kwiat,yamamoto,pushin}.
In our case, we have to protect signals from spatially separated independent senders, thus
our codes have to be factorizable according to this separation.
It turns out that in the considered setting the existence of such codes is exactly equivalent to the
 certain decomposability properties in the space of matrices of bounded rank. One of the main
  motivations for the present paper was merging these concepts in a fruitful way.

  It is worth noting that our problem is
equivalent to finding a product subspace in a given subspace. One
may think of this problem as of the generalization of the well-known (solved) 
problem of the existence of a product state in a
given subspace \cite{hindus,product-local}.

The paper is organized as follows. In Sect.~2, we provide some background material. In Sect.~3,
we state our main results. Next, we apply the results to the specific case of a qutrit-qutrit input. Then, we discuss some generalizations.

\section{Background}

In this section we give some necessary background material. This includes quantum error
 correction conditions, spaces of matrices of bounded rank, and several miscellaneous mathematical facts. For the reader's convenience this part is quite voluminous.

\subsection{Quantum channels}

Quantum channel $\calL$ is a completely positive trace-preserving map. Every channel
admits the so-called Kraus (or Choi--Kraus, or operator-sum) representation as follows
$\mathcal{L}(\varrho)=\sum_i A_i \varrho A_i^{\dagger}$ with $\sum_i A_i^{\dagger}A_i=\mathbbm{1}$ 
\cite{Choi,Kraus}.
 A random unitary channel has the representation
 \beq\label{ruc}
 \calL(\varrho)\;=\;\sum_i p_i U_i \varrho U_i^{\dagger}\,,
 \eeq
  where $U_i$
  are unitary and $\sum_i p_i =1$, $p_i\ge 0$. When such a channel has two Kraus operators,
  \tzn, $\calL(\varrho)=pU_1\varrho\dager{U_1}+(1-p)U_2\varrho\dager{U_2}$, 
  it is called bi-unitary (or binary unitary; shortly BUC).
  This kind of channels will soon be the starting point of our considerations.

Channels can be classified upon the number of senders and receivers using them.
 We have the  following types of channels according to such a classification 
 \cite{BKN,md-ph-1,yard,md-ph-2,broadcast}:
\begin{itemize}
\item bipartite --- one sender and a single receiver,
\item multiple access --- several senders and a single receiver,
\item broadcast --- a single sender and several receivers,
\item $km$-user --- $k$ senders and $m$ receivers, $k,m>1$ (if $k=m$ then we have a $k$-user channel)
\end{itemize}

\begin{figure}[t]
\centering
\includegraphics[width=8cm,height=6.5cm]{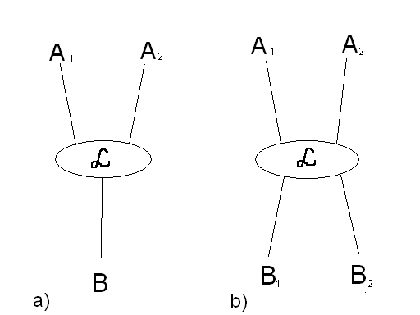} 
\caption{\label{rysunek} The geometrical structure
of (a) a  two-access channel and (b) a $2$-user channel. In the latter case we will be interested in the communication $A_i\rightarrow B_i$. }
\end{figure}

Multiple access channels (MACs) and $k$-user channels ($k$-UCs)
are the main scope of this paper. For $k$-UC we will be
interested in a reduced communication, which means a scenario in
which each sender wants to transmit quantum information to only
one receiver ($i$-th sender to $i$-th receiver; see Fig.~\ref{rysunek}).

Due to the possibility of a rotation $U_i^{\dagger}(\,\cdot\,)U_i$ on the
output or the input of a channel, in bipartite, multiple access
and broadcast case one can consider a simplified random unitary
channel in general reasonings\footnote{This is also true for a
general $km$-user channel if one of the unitaries $U_i$ is
product across the cuts corresponding to the separation of either
senders or receivers. Naturally, in a concrete case one needs to
remember that $\tilde{U}_k=\dager{U_i}U_k$ or $U_k\dager{U_i}$.}
\beq\label{rotowany} 
\tilde{\calL}(\varrho)\;=\;p\varrho+\sum_{k} p_k
\tilde{U}_k\varrho\dager{\tilde{U}_k}  \eeq
with $\sum_i p_i=1-p$.
For a
MAC (with $k$ inputs) and a $k$-UC it holds
$\varrho=\bigotimes_{i=1}^k\varrho_i$, where $\varrho_i$ is the
input of the $i$-th sender.

\subsection{Quantum error correction}\label{qec}

QECC is a subspace $\calC$ of a larger Hilbert space $\calH$. Equivalently, a code is defined
to be the projection $P_{\calC}$ onto $\calC\subseteq\calH$. One says that $\calC$ is correctable
 if all states $\varrho=P_{\calC}\varrho P_{\calC}$ can be recovered after action of a channel using
  some decoding operation $\calD$, that is $\calD\circ\calL (\varrho)=\varrho$. Such recovery
  operation exists if and only if $P_\calC  A_i^{\dagger}A_j P_\calC=\alpha_{ij}P_\calC$ for some
  hermitian matrix $L=[\alpha_{ij}]$. These conditions are due to Knill and Laflamme (KL) \cite{KL}.
   In case of many usages of a channel, $A_i$ are tensor products of Kraus operators in KL conditions.
   In this paper, however, we concentrate on a single usage of a channel.

When we have a larger number of senders we talk about a local code
$\calC_i$ assigned to each of the senders. It is an immediate
observation that KL conditions need only a little adjustment to
serve for the case of MACs. Namely, we have (with the obvious
notation):
\begin{rem}
Local codes $\calC_i$ are correctable for a MAC with Kraus operators $\{A_i\}$ with $k$ inputs if and only if
\beq\label{KL-MAC} P_{\calC_1}\otimes P_{\calC_2}\otimes\dots\otimes P_{\calC_k}  A_i^{\dagger}A_j P_{\calC_1}
\otimes\dots\otimes P_{\calC_k}\;=\;\alpha_{ij}P_{\calC_1}\otimes\dots\otimes P_{\calC_k}
\eeq
 for some hermitian matrix $L=[\alpha_{ij}]$.
\end{rem}
This is true since the set of product codes is a subset of the set
of all codes.  For a larger number of receivers the situation is
more involved as one requires also from decoding to be product
across proper cuts.

 In further parts, we use the notation
$R\otimes R'$ or $S\otimes S'$ for a code for a channel with two
inputs and talk about $M\otimes N$ codes, where $M,N$ denote
the dimensions of local codes.

In some cases communication with a given code is free of
decoherence, which means that either the output of a channel is
already the same as the input or the output needs only a unitary
operation to be recovered to the initial state. In the former case
we talk about {\it decoherence-free subspaces} (DFSs), in the
latter about {\it unitarily correctable codes} (UCCs)
\cite{kribs-pasieka-entropy-code}. Both correspond to $L$ being a
pure state, \tzn, $\alpha_{ij}=\alpha^*_i \alpha_j$ holds and for
this reason they are also called the zero entropy codes
\cite{kribs-pasieka-entropy-code}.
 Clearly all DFSs are UCCs.
 In further parts we will use a general name decoherence-free 
 codes (DFC) for them. They are the main interest of the
present paper. Due to our interest in channels with multiple
accesses we wish to find DFCs which factorize according to the
spatial separation of senders.

\subsection{Spaces of matrices of bounded rank}
\label{spaces-matrices}

A space of matrices of bounded rank is a space which contains only elements whose ranks are bounded by some prescribed number.
If a space contains only elements of rank equal to $k$ (naturally, besides the zero element) we talk about $k$-spaces.
The research on spaces of matrices of bounded (equal) rank dates back to works by Flanders \cite{flanders} and Westwick \cite{westwick}.
In quantum information theory the concept of spaces of matrices of bounded rank were recently used in \cite{cubitt}.

We will use $\frakA, \frakB, \frakC, \ldots$ to denote space of matrices spanned respectively by $A_i, B_i, C_i, \ldots$.
We define $r_m(\frakA)\equiv \max_{A\in \frakA} r(A)$.
Further, we recall the concept of the equivalence of spaces of matrices \cite{flanders} and the decomposability  \cite{westwick}.
We say that $\frakA$ is {\it equivalent} to $\frakB$ if there exist nonsingular matrices $E$,$F$ such that
$\frakA=\{EBF,\;B\in\frakB\}$. If these matrices are explicitly specified we talk about {\it $(E,F)$-equivalency}.
A subspace $\frakA$ of $a\times b$ matrices is called {\it $(t,s)$-decomposable} if it is equivalent to a subspace whose all elements have the form
\beq\label{decomp} 
A=\left[
  \begin{array}{cc}
    [0]_{(a-t)\times (b-s)} & B_{(a-t)\times s} \\
    C_{t \times (b-s)} & D_{t\times s}\\
  \end{array}
\right],\quad A\in\frakA\,.
\eeq
 When $t,s$ are not specified, we will be just talking about a general fact of {\it $(t+s)$-decomposability}.
  Below we collect several important  facts concerning decomposability in the spaces of $3\times 3$ matrices of
   bounded rank. It holds true that [29,\,31\,--\,33]
 \begin{itemize}
 \item three-dimensional $2$-subspaces are not $2$-decomposable; moreover, up to an equivalence,
  there exists only one such space --- the space of skew-symmetric matrices\footnote{It is an interesting coincidence:
   this space corresponds to the antisymmetric space of two qutrits  via the identification states-matrices; antisymmetric
   spaces play an important role in many applications of quantum information theory.},
 \item if $\frakB$ is a $2$-subspace then it necessarily holds that $\dim\frakB\le 3 $,
 \item if a subspace $\frakB$ with $r_m(\frakB)=2$ contains a rank one matrix then it is $2$-decomposable,
 \item a subspace $\frakB$ with $r_m(\frakB)=2$ and $4\le \dim \frakB \le 6$ is $2$-decomposable (follows from the above).
 \end{itemize}

\subsection{Space of states vs.~space of Schmidt matrices}\label{versus}

 The well known one-to-one identification of pure states with matrices will be
one of our main tools:
with fixed orthonormal bases $\ket{i}$ and $\ket{j}$ for $\calH_1=\mathbb{C}^{d_1}$ and $\calH_2=\mathbb{C}^{d_2}$
 respectively, one defines the {\it Schmidt matrix} of a state $\ket{\psi}=\sum_{ij}c_{ij}\ket{ij}$ to be $C=\sum_{ij}c_{ij}\outerp{i}{j}$.
For two states $\ket{\phi}$ and $\ket{\psi}$ with Schmidt matrices $C$ and $D$ it holds $\inner{\phi}{\psi}=\mathrm{tr} \dager{C}D$.
By direct calculation one finds that the transformation $\ket{\psi} \mapsto U_1\otimes U_2 \ket{\psi}$ corresponds to  $C\mapsto U_1 C U_2 ^T$.
Further,
we define the {\it Schmidt rank} $r$ of a state $\ket{\psi}$, denoted by $r(\ket{\psi})$, to be the number of nonzero elements in its Schmidt decomposition and
the {\it maximal Schmidt rank} $r_{m}$ of the subspace $\calH$ to be
$r_{m} (\calH)\equiv \max_{\ket{\psi}\in\calH} r(\ket{\psi})$.
 Obviously $r(\ket{\psi})=r(C)$.

Let now $\calH=\spann\{\ket{\gamma_i}\}$ and $\frakH=\spann\{h_i\}$, where $h_i$ is a 
Schmidt matrix of $\gamma_i$.
Then $r_m(\calH)=r_m(\frakH)$ due to the isomorphism between the space of unnormalized states and matrices. It is
 thus natural to transfer the idea of decomposability of the space of matrices to the space of states. Therefore we propose to use the following:
\begin{defi}
A subspace $\calH$ is called $(i,j)$-decomposable if $\frakH$ is so.
\end{defi}

\section{Main Results}
\label{main-results}

As announced earlier, we begin with bi-unitary channels, for
which (\ref{rotowany}) turns into
\beq\label{buc}
\tilde{\calL}(\varrho)\;=\;p\varrho+(1-p)U\varrho\dager{U}\,. 
\eeq
For such channels the set of DFS and UCC coincide. This means that
a DFC constructed for a MAC will also work fine for a reduced
communication through a $k$-UC.

For one use of a BUC, KL
conditions (\ref{KL-MAC}) reduce to a {\it single}
 condition $PUP=\lambda P$ which for MACs takes the form\footnote{\label{uwaga-o-rankach} Motivated by the form of KL
 conditions, the authors of \cite{choi-et-al} introduced the notion of the higher rank numerical range as the tool helpful in finding QECs for bipartite channels. For an arbitrary operator $A$, the higher rank (or rank-$k$) numerical range is the set $\Lambda_k(A)=\{\lambda\in \mathbb{C}:PAP=\lambda P\;\;\mathrm{for\;some}\;\;P\in\calP_k\}$, where $\calP_k$ is the set of all rank $k$ projections. This notion and the form of KL conditions for MACs prompted us to introduce the notion of the {\it $k_1\otimes k_2\otimes\cdots$ product higher rank numerical range}, which for an operator $A$ is defined to be $\Lambda_{k_1\otimes k_2\otimes\cdots}(A)=\{\lambda\in\mathbb{C}:R\otimes R'\otimes\cdots AR\otimes R'\otimes\cdots=\lambda R\otimes R'\otimes\cdots\;\;\mathrm{for\;some}\;\; R\in \calP_{k_1},\;R'\in\calP_{k_2},\ldots\}$. Properties of this set are considered in \cite{my}.}
\beq\label{condition-zwykly} 
R\otimes R'UR\otimes R'\;=\;\lambda
R\otimes R'\,. 
\eeq

Let us consider the case of two $d$-dimensional inputs (and thus a
$d^2$-dimensional output) and take $U$ to be hermitian. The latter
implies that
\beq \label{uuu} U=P-Q\,, \eeq
where $P$ and $Q$ are
both projections.
We further denote the eigensubspaces of $U$ as $\calP=P\calH$ and $\calQ=Q\calH$ and 
let $\dim \calP=p$ and  $\dim
\calQ=q$. For obvious reasons, we assume $pq>0$ to hold.

We have the following
concerning the channel from (\ref{buc}) with $U$ taken as in (\ref{uuu}):
\begin{thm}\label{glowne-tw}
An $M\otimes N$ DFC exists if and only if at least one of the
subspaces $\calP$ or $\calQ$ is  $(d-M,d-N)$-decomposable.
\end{thm}
\begin{proof}
We put $\lambda=\pm 1$ in (\ref{condition-zwykly}) as
these are the suspicious values ($L$ is then pure). Using (\ref{uuu}) we have:
\begin{equation}\label{plus}
R\otimes R' (P-Q) R\otimes R'\;=\; R\otimes R'\,, \quad \lambda =1
\end{equation}
and
\begin{equation}\label{minus}
S\otimes S' (P-Q) R\otimes R'\;=\; - S\otimes S'\,, \quad \lambda =-1\,,
\end{equation}
where $R,S$ and $R',S'$ are projections rank $M$ and $N$,
respectively. It is evident that it is sufficient to conduct
calculation for only one of the equations above as the
 sign of
$\lambda$ exchanges only the roles of $P$ and $Q$. Therefore we
concentrate on (\ref{plus}), which, due to $P+Q=\mathbbm{1}$,
can be rewritten in two {\it equivalent} forms as
\begin{equation}\label{zero-plus}
R\otimes R' Q R\otimes R'\;=\;0
\end{equation}
and
\begin{equation}\label{zero-plus-bis}
R\otimes R' P R\otimes R'\;=\;R\otimes R'\,.
\end{equation}
Let us concentrate on the first equation (further, in some cases, the second one will be  more convenient for drawing conclusions).
Assume
\begin{equation}\label{pe}
Q\;=\;\sum_{i=1}^q \proj{\phi_i}\,,\quad 
\ket{\phi_i}\;=\;\sum_{kl} c_{kl}^{(i)} \ket{kl}\,,\quad k,l=1,\ldots,d
\end{equation}
and let $[C_i]_{kl}=c^{(i)}_{kl}$.
We can represent the projections as rotated projections written in the standard basis as
\begin{equation}\label{rozklad1}
R\otimes R' \;=\;U_1^{\dagger}\otimes U_2^{\dagger} \tilde{R}\otimes \tilde{R}' U_1\otimes U_2\,,
\quad \tilde{R}\;=\;\sum_{g=0}^{M-1}\proj{g}\,,\quad \tilde{R}'\;=\;\sum_{h=0}^{N-1}\proj{h}\,,
\end{equation}
where $U_1$ and $U_2$ are unitary. Inserting (\ref{pe}) and (\ref{rozklad1}) into  
(\ref{zero-plus})
and taking into account that a matrix is zero iff all its elements are so we arrive at
\begin{eqnarray}\label{jeden}
\bra{ij} \Big(U_1\otimes U_2 \Big(\sum_{m} \proj{\phi _m}\Big)
U_1^{\dagger}\otimes U_2^{\dagger}\Big) \ket{kl}\;=\;0\,,\nonumber\\
\quad i,k=0,1, \dots, M-1\,,\qquad j,l=0,1,\dots, N-1\,.
\end{eqnarray}
In particular, this must be true for $ij=kl$, which gives
\begin{eqnarray}
&\sum_{m}|\bra{ij} U_1\otimes U_2 \ket{\phi _m}|^2
\;=\;0\,,& \nonumber\\
& i=0,1,\ldots,M-1\,, \qquad j=0,1,\ldots,N-1 &
\end{eqnarray}
and consequently for {\it all} values of $m$
\begin{eqnarray}\label{glowne}
&\bra{ij}U_1\otimes U_2 \ket{\phi_{m}}\;=\;0\,, & \nonumber\\ 
& i=0,1,\ldots,M-1\,,
\qquad j=0,1,\ldots,N-1\,. &
\end{eqnarray}
This condition implies that off-diagonal terms vanish as well and thus 
(\ref{glowne}) is fully equivalent to (\ref{zero-plus}). Recalling the transformation rule for Schmidt matrices under local unitary
rotations of a state (see Sect.~2.4) condition above can be rewritten as
\begin{eqnarray}\label{transformacja}
&\forall\; m \quad \bra{i}U_1 C_{m} U_2 ^T \ket{j}\;=\;0\,, & \nonumber\\ 
& i=0,1,\ldots, M-1\,,\quad j=0,1,\ldots, N-1\,. &
\end{eqnarray}
Denote $U\equiv U_1$ and $V\equiv U_2^T$. Equation
(\ref{transformacja}) states that all Schmidt matrices $C_i$ must
brought with the same unitaries $U,V$ to a form with the $M\times
N$ zero matrix in the upper left corner. Notice that instead of
full unitary matrices $U,V$ we can consider their reductions,
\tzn, isometries $U_{\rm isom}$ and $V_{\rm isom}$, which are $M\times d$
and $d\times N$, respectively. Further, we need the following
simple lemma:
\begin{lem}
Let $A_i$ be complex  matrices and let
 $V_1$, $V_2$ be isometries. The condition $V_1 A_i V_2=[0]_{M\times N}$ holds for all 
 values of an index
 $i$ if and only if  for all complex $\vec{\alpha}=(\alpha_1,\alpha_2,\ldots)$ it holds 
 that $V_1 (\sum_{i}\alpha_i C_i) V_2=[0]_{M\times N}$.
\end{lem}
This means that the space $\frakC$ must be 
$(U,V)$-equivalent to a space whose all elements have the
zero $M\times N$ matrix in the upper left corner. It suffices now to show that this equivalency means also $(d-M,d-N)$-decomposability,
 \tzn, to show that unitary matrices are as powerful as general full rank matrices in definition of decomposability (\ref{decomp}).
This is what we are going to demonstrate now.  Let $X$, $Y$ be nonsingular matrices.
If $XC_iY$ has an $M\times N$ zero block it means that 
$X=([\tilde{X}^T]_{d\times M},[X']_{d\times(d-M)})^T$
and $Y=([\tilde{Y}]_{d\times N},[Y']_{d\times (d-N)})$, where  $\tilde{X}$ and $\tilde{Y}$ are respectively
rank $M$ and $N$ matrices such that $\tilde{X}C_i\tilde{Y}=[0]_{M\times N}$.
It means
that $\tilde{X}C_i$ has rank less than or equal to $d-N$ and the (possibly nonorthogonal) rows of $\tilde{Y}$
span the remaining $N$ dimensions allowed in a $d$ dimensional space. Thus it is enough to take isometry whose
rows span the same space as the ones of $\tilde{Y}$ do to achieve zeroing of the resulting matrix. Arguing in
the same way for the left multiplication we arrive at the conclusion that both $\tilde{X}$ and $\tilde{Y}$ can
 be replaced by isometries, which can be further completed to unitaries. This ends the proof of the theorem. 
\end{proof}%
Let now $\frakD$ be the space spanned by Schmidt matrices of the
projection $P$. We can immediately conclude the following simple
necessary condition for the existence of a DFC:
\begin{cor}\label{konieczny}
If there is an $M\otimes N$ DFC then either of the following holds
\begin{eqnarray} 
r_m(\frakC) &\le& 2d-(M+N)\,, \\
r_m(\frakD) &\le&
2d-(M+N)\,.
\end{eqnarray} 
Alternatively 
\begin{eqnarray} 
r_m(\calQ) &\le& 2d-(M+N)\,,\\
r_m(\calP) &\le& 2d-(M+N)\,.
\end{eqnarray}
\end{cor}
\begin{proof}
Follows from Theorem \ref{glowne-tw} and the property 
\[ r(B)\;\le\; t+u -r(A)-r(C)+r(ABC)
\] 
valid for
arbitrary matrices $A_{s\times t}$, $B_{t\times u}$, $C_{u\times v}$. The latter can be proved by using
  twice the Sylvester's inequality \cite{lutkepohl-handbook}.
\end{proof}

For further practical purposes, it is convenient to rewrite Theorem \ref{glowne-tw} as follows:
\begin{cor}\label{cor}
An $M\otimes N$ DFC exists if and only if either of the following
holds
\begin{enumerate}
\item there exists an isometry $[V_1^Q]_{M\times d}$ such
that 
\beqn r\left( \calC _V^Q := \left[
                               \begin{array}{c}
                                 V_1^Q C_1  \\
                                 V_1^Q C_2 \\
                                 \cdots \\
                                 V_1^Q C_q
                               \end{array}
                             \right]
   \right)\;\le\; d-N\,,
   \eeqn
\item there exists an isometry $[V_1^P]_{M\times d}$ such that
\beqn
r\left( \calC _V^P := \left[
                               \begin{array}{c}
                                 V_1^P D_1  \\
                                 V_1^P D_2 \\
                                 \cdots \\
                                 V_1^P D_p
                               \end{array}
                             \right]
   \right)\;\le\; d-N\,.
   \eeqn
\end{enumerate}
\end{cor}

We observe  that we can give immediately a rough bound on
dimensions of an input for which an $M\otimes N$ code surely exists
for a fixed $q$ (naturally, the same will hold with $p$ in place
of $q$).
\begin{cor}\label{wystarczajacy}
If
\begin{enumerate}
\item[i)] $q=2$ and $d\ge M+N$
or 
\item[ii)] $q>3$ and $d \ge \min\{ Mq +N, Nq+M\}$ 
\end{enumerate}
then there always exists an $M\otimes N$ DFC for the considered channel.
\end{cor}
\begin{proof}
i) follows from the generalized Schur decomposition theorem \cite{stewart} stating that
for any two complex matrices $A$, $B$ there exist unitary transformations $U$, $V$ such that $UAV^{\dagger}$ and $UBV^{\dagger}$
are both upper or lower triangular at the same time.

As to ii).
$M$ rows of $q$ Schmidt matrices of $Q$ span at most $Mq$ dimensional subspace. Since $d\ge Mq+N$ there is
still enough space for $N$ orthogonal vectors in the whole $d$ dimensional space. Taking $U_{\rm isom}=(\mathbbm{1}_{M\times (d-M)}[0]_{M\times M})$ and $V_{\rm isom}$ to
 consist of the mentioned $N$ vectors as matrix columns we produce the $M\times N$ zero matrix. Analogous reasoning applies to
 the right multiplication by $V_{\rm isom}$ and choosing $U_{\rm isom}$ to consist of proper columns.
\end{proof}%

Clearly, the results of this section can also be formulated  for a
general $d_1 \otimes d_2$ input (and thus a $d_1\times  d_2$
output).

\section{Applications: Qutrit-Qutrit Input Case}
We now move to the specific case of qutrit inputs to a channel.
Our main interest will be in the construction of $2\otimes 2$
codes. The cases $1\otimes 2$ and $2\otimes 1$ are not interesting
for us as
 such codes correspond to a nonzero rate only on the single wire of the network, on the other hand, codes $2\otimes 3$ and $3\otimes 2$ correspond to the situations when one of the parties can send with the maximum rate, which clearly requires very specific type of noise (although not of the form $U'\otimes \mathbbm{1}$ or $\mathbbm{1}\otimes U'$ as one can check).

\subsection{General approach}

Direct application of Theorem \ref{glowne-tw} relying on finding proper isometries $U_{\rm isom}$ and $V_{\rm isom}$
giving zero blocks in the corner of all the matrices
is a tedious task when $3\le q\le 7$. In these cases, one is thus recommended to consult 
[29,\,31\,--\,33] for elegant techniques. Some relevant results have been given in Sect.~\ref{spaces-matrices}

In what follows we concentrate on the $q=2$ case. In \cite{as-rank2}, it has been shown that every two
dimensional spaces of matrices with $r_m\le k$ is $k$-decomposable. However, no explicit distinction between different
kinds of decomposability has been given. We thus feel inclined to provide detailed discussion concerning $(1,1)$-decomposability in the
relevant case of $d=3$. The result is summarized in the following theorem:
\begin{thm}\label{centralne}
If\, $q=2$ then a $2\otimes 2$ DFC exists if and only if
$r_m(\frakC)\le 2$ \emph{and} $\frakC$ is \emph{not} a
$(0,2)$ or $(2,0)$-decomposable $2$-subspace.
\end{thm}
\begin{proof}
(The proof is constructive) It is obvious that we need only to solve (\ref{plus}),
 that is to take $\lambda=+1$. 
Thus, no considerations concerning decomposability of $\frakD$ are required.
  Further, it is also evident from Corollary \ref{konieczny} that it must be that $r_m(\frakC)\le 2$ implying
  that $r(C_i)\le 2$. We assume that we have already passed to the $(W_1,W_2)$-equivalent space where $W_1$ and $W_2$
   stem from the singular value decomposition of $C_2$. In such a basis we assume these matrices to be
\begin{equation} 
C_1\;=\;\left[
       \begin{array}{ccc}
         c_{11} &c_{12}  & c_{13} \\
         c_{21} & c_{22} & c_{23} \\
         c_{31} & c_{32} & c_{33} \\
       \end{array}
     \right], \quad C_2\;=\;\left[
        \begin{array}{ccc}
          0 & 0 & 0 \\
          0 & b & 0 \\
          0 & 0 & a \\
        \end{array}
      \right],\quad a+b> 0\,.
      \end{equation}
It is further assumed that $r(C_1)\le r(C_2)$.

Suppose $r_m(\frakC)=1$. It is a simple observation that in this case, only the last row or the last column
(not at the same time obviously) 
in the matrix $C_1$  is nonzero. Existence of a code is thus a trivial fact.

Suppose now that $\frakC$ contains an element of rank $2$. W.l.o.g. we can assume that $C_2$ is such an element, that is both $a$ and $b$ are greater than zero.
 Condition $r_m(\frakC)\le 2$ is fulfilled if for all $\beta$ it holds that $\det (C_1+\beta C_2)=0$,
 which happens when\footnote{Conditions of this type are standard in analyses of spaces of matrices of bounded rank.}
\beq\label{war1}
c_{11}=0\,,\quad \det C_1=0\,,\quad a c_{12} c_{21}+b c_{13} c_{31}=0\,,
\eeq
which we hereafter assume to hold.
We now further consider specific cases.

Suppose $(c_{12},c_{13})\ne (0,0)$ and $(c_{21},c_{31})\ne (0,0)$.
Take 
\beqn\label{oblozenia} 
V_1^Q\;=\;\left[
        \begin{array}{ccc}
          1 & 0 & 0 \\
          0 & \displaystyle\frac{ac_{12}}{N_1} & \displaystyle\frac{bc_{13}}{N_1} \\
        \end{array}
      \right],\qquad
V_2^Q\;=\;\left[
               \begin{array}{cc}
                 1 & 0 \\
                 0 & \displaystyle\frac{ac_{21}}{N_2} \\
                 0 & \displaystyle\frac{bc_{31}}{N_2} \\
               \end{array}
             \right],\\
N_1\;=\;\sqrt{a^2|c_{12}|^2+b^2|c_{13}|^2}\,, \qquad
N_2\;=\;\sqrt{a^2|c_{21}|^2+b^2|c_{31}|^2}\,. 
\eeqn
 With this choice of isometries, using the last condition of 
 (\ref{war1}), one can verify that it holds for the
 matrix elements that $[V_1^Q C_1 V_2^Q]_{11}=[V_1^Q C_1 V_2^Q]_{12}=[V_1^Q C_1 V_2^Q]_{21}=0$ and $V_1^Q C_2 V_2^Q=[0]_{2\times 2}$;
  using the last two conditions of (\ref{war1}) one arrives at $[V_1^Q C_1 V_2^Q]_{22}=0$. This proves the existence of a code.

If $(c_{12},c_{13})\equiv(0,0)$ and $(c_{21},c_{31})\equiv(0,0)$  by the previously mentioned generalized Schur
decomposition and a simple swapping of rows we can transform both matrices simultaneously according to
\beq
\left[
  \begin{array}{ccc}
    0 & 0 & 0 \\
    0 & \star & \star  \\
    0 & \star & \star \\
  \end{array}
\right]\;\longrightarrow\;
\left[
  \begin{array}{ccc}
    0 & 0 & 0 \\
    0 & \star & \star  \\
    0 & 0 & \star \\
  \end{array}
\right]\;\longrightarrow\;
\left[
  \begin{array}{ccc}
    0 & 0 & 0 \\
    0 & 0 & \star  \\
    0 & \star & \star \\
  \end{array}
\right],
\eeq
which outputs the proper code.

Suppose $(c_{12},c_{13})\equiv(0,0)$ and $(c_{21},c_{31})\neq (0,0)$. Setting $[V_1^Q]_{ij}=v_{ij}$ we have
\begin{equation}
\calC_V^Q\;=\;\left[
          \begin{array}{ccc}
            v_{12}c_{21}+v_{13} c_{31} &v_{12} c_{22}+v_{13} c_{32}& v_{12} c_{23}+v_{13} c_{33}\\
            v_{22}c_{21}+v_{23} c_{31} & v_{22} c_{22}+v_{23} c_{32}& v_{22} c_{23}+v_{13} c_{33} \\
            0 & bv_{12} & av_{13} \\
            0 & bv_{22} & av_{23} \\
          \end{array}
        \right],
\end{equation}
which must be rank one (all rows, equivalently columns, must be proportional to each other).
 There are two (in principle nonexclusive) possibilities for this to hold: (i) $bv_{12}=av_{13}=0$ and  $bv_{22}=av_{23}=0$ or (ii) $v_{12}c_{21}+v_{13} c_{31}=0$ and $v_{22}c_{21}+v_{23} c_{31}=0$.
The first alternative cannot be true in any case since $a>0$, $b>0$ and this would entail the fact that $v_{12}=v_{13}=v_{22}=v_{23}=0$
which is impossible because of the isometric character of $V_1^Q$. We thus have
$v_{12}c_{21}+v_{13}c_{31}=0$ and $v_{22}c_{21}+v_{23} c_{31}=0$. W.l.o.g. we can set $v_{22}=v_{23}=0$ reducing the
problem to finding conditions under which there exists such $\gamma$ that the system of equations
\begin{equation}
 \begin{array}{rcl}
  v_{12}c_{21}+v_{13}c_{31}&=&  0\,,  \\[1ex]
  v_{12} c_{22}+v_{13} c_{32}& =& \gamma b v_{12}\,, \\[1ex]  
  v_{12} c_{23}+v_{13} c_{33}&=& \gamma a v_{13}\,,
\end{array}
\end{equation}
has a nontrivial solution, where the last two equations are the requirement that the first row is
 proportional to the third one. This is possible if there exists such $\gamma$ that
\beqn 
r\left(\left[
    \begin{array}{ccc}
               c_{21} & c_{22}-\gamma b & c_{23} \\
               c_{31} & c_{32} & c_{33}-\gamma a \\
                                                  \end{array}
                                                \right]
\right)\;\le\; 1\,.
\eeqn
Recalling that the first row of $C_1$ is now equal to zero, it can be equivalently written  as $r(C_1-\gamma C_2)\le 1$,
 which is simply the obligation for the subspace $\calQ$ to contain a rank one element (there must {\it exist} such $\gamma$,
 not for all of them it must hold) or, in other words, {\it not} to be a $2$-subspace. The case of $(c_{12},c_{13})\ne(0,0)$
  and $(c_{21},c_{31})= (0,0)$ can be treated obviously in a similar manner and we get that there must exist such $\gamma$ that
\beqn 
r\left(\left[
                                                  \begin{array}{cc}
                                                    c_{12} & c_{13} \\
                                                    c_{22}-\gamma b & c_{23}\\
                                                    c_{32}&  c_{33}-\gamma a \\
                                                  \end{array}
                                                \right]
\right)\;\le\; 1
\eeqn
with the same conclusion.
This  exhausts all possibilities. 
\end{proof}%
With no effort we can extend our result to the case $q=7$. Namely, we have
\begin{thm}
If $q=7$ then a $2\otimes 2$ DFC exists if and only if
$r_m(\frakD)\le 2$ \emph{and} $\frakD$ is \emph{not} a
$(0,2)$ or $(2,0)$-decomposable $2$-subspace.
\end{thm}%
Notice that at no point in the proof have we made the assumption about orthogonality of the 
matrices.

 We conclude this section with some observations.
\begin{rem}\label{produktowosc}
Let $P$, $Q$, $R$ be projectors with $r(P)=p$, $r(Q)=q$, $r(R)=pq$. The following holds
$P\otimes Q R P\otimes Q=P\otimes Q$ if and only if $R=P\otimes Q$.
\end{rem}

This means that in some situations we can approach the problem of deciding the existence of a code more directly for $q=4,5$.
Specifically, for $q=4$ we check whether $Q$ is product $2\otimes 2$. Positive answer resolves the matter on the spot. If the answer is negative, we need to check $\frakD$ for the decomposability.  By analogy, if $q=5$ we check whether $P$ is  $2\otimes 2$ product, if it is we immediately have a code, if not --- we test $\frakC$ for decomposability. Using this method we can easily show that $\lambda=-1$ and $\lambda=+1$ 
cannot be solution to (6) at the same time. We call this impossibility the {\it uniqueness} of a DFC. We state this fact as follows:
\begin{rem}\label{jedyny}
A $2\otimes 2$ DFC for the considered noise model in case of $d=3$
is unique.
\end{rem}%
Interestingly, the uniqueness is more powerful as the following
holds true (for the proof see \cite{my}):
\begin{rem}
If there is a $2\otimes 2$ DFC for $d=3$ there exists no code
corresponding to other values of $\lambda$.
\end{rem}

\subsection{Examples}

Here we provide several illustrations to the results obtained
above. In what follows, we assume (\ref{pe}) to hold; in every
case our concern is in the existence of a $2\otimes 2$ code.  We
itemize our examples below:
\begin{itemize}
  \item $q=1$,
  \[
  \ket{\phi_1}\;=\;\frac{1}{\sqrt{3}}(\ket{00}+\ket{11}+\ket{22})\,.
  \]
  A code does not exist since the necessary condition (Corollary \ref{konieczny}) is not fulfilled as $r(C)=3$.
  \item $q=2$,
  \[
  \ket{\phi_1} \;=\; \frac{1}{\sqrt{2}}(\ket{11}+\ket{22})\,,\qquad
  \ket{\phi_2} \;=\; \frac{1}{\sqrt{2}}(\ket{10}+\ket{21})\,.
  \]
  $\frakC$ is a $(0,2)$-decomposable $2$-subspace and as such it is not $(1,1)$-decomposable 
  thus a code does not exist (see Theorem \ref{centralne}).
  \item $q=2$,
    \[
  \ket{\phi_1}\;=\;\frac{1}{\sqrt{2}}(\ket{02}+\ket{10})\,,\qquad
  \ket{\phi_2}\;=\;\frac{1}{\sqrt{2}}(\ket{01}+\ket{20})\,.
  \]
  $\frakC$ is a $2$-subspace but its $(1,1)$-decomposability can be easily seen. 
  A code exists and is given by $P_C=R\otimes R'=P_{12}\otimes P_{12}$, where $P_{12}=\proj{1}+\proj{2}$.
  \item $q=3$,
   \begin{eqnarray*}
  & \ket{\phi_1}\;=\;\frac{1}{\sqrt{2}}(\ket{01}-\ket{10})\,,\qquad
  \ket{\phi_2}\;=\;\frac{1}{\sqrt{2}}(\ket{12}-\ket{21})\,, & \\
  &
  \ket{\phi_3}\;=\;\frac{1}{\sqrt{2}}(\ket{02}-\ket{20})\,.&
  \end{eqnarray*}
  $Q$ is a projection onto the antisymmetric subspace, thus a DFC does not exist. This example is of particular importance since the projection corresponds to
  the noise with $U$ being a SWAP gate. In fact, there is also no other code for  two-access transmission through such a channel (see \cite{my}).
  \item $q=4$
   \[
  \ket{\phi_1}\;=\;\ket{00}\,,\quad
  \ket{\phi_2}\;=\;\ket{01}\,,\quad
  \ket{\phi_3}\;=\;\ket{10}\,,\quad
  \ket{\phi_4}\;=\;\ket{11}\,.
  \]
  $Q$ is product itself so a code definitely exists. Alternatively, we could check $\frakC$ and $\frakD$ which will result in $(1,1)$-decomposability of $\frakD$.
  \item $q=4$
  \[
   \ket{\phi_1}\;=\;\frac{1}{\sqrt{2}}(\ket{00}+\ket{11})\,,\quad
  \ket{\phi_2}\;=\;\ket{20}\,,\quad
  \ket{\phi_3}\;=\;\ket{21}\,,\quad
    \ket{\phi_4}\;=\;\ket{22}\,.
  \]
 It holds that $r_m(\calQ)=3$ so $\calQ$ cannot be $(1,1)$--decomposable. On the other hand, $Q$ itself is not product so there is no code at all. It is interesting that $\calQ$ contains a rank three vector $1/\sqrt{3}(\ket{00}+\ket{11}+\ket{22})$ but its complement contains vectors of rank 
 at most two.\footnote{The example has been suggested to the author by P. Horodecki.} 
The latter
 leads to:
 \begin{rem}
 In a $3\otimes 3$ Hilbert space, the complement $\calP$ of the four dimensional subspace $\calQ$ with $r_m(\calQ)=3$ may have $r_m(\calP)=2$.
 \end{rem}
\end{itemize}

\section{Generalizations}

In this section we discuss several generalizations of the results presented in the previous sections.

Consider a unitary operation:
\beq\label{generic}
U\;=\;\sum_k \eksp^{\uroj \delta_k}P_k
\eeq
with some arbitrary phases $\delta_k$ (one phase, say $\delta_0$, can always be taken equal to zero) and projections $P_k$.
An $M\otimes N$ (in general $MN$-dimensional) DFC can exist iff there exists an eigenvalue with degeneracy of order at least $MN$.
 It follows that our technique of finding product subspaces is also applicable in a generic case of (\ref{generic}). Indeed, assume
  for simplicity $z=1$ is the eigenvalue with at least $MN$-fold degeneracy (and only this highly degenerate eigenvalue), then $U=P_0+\sum_{k\ne 0} e^{i\delta_k}P_k$, with $r(P_0)\ge MN$ and $\delta_k\ne0$. We need to solve the KL condition $R\otimes R' U R\otimes R'=R\otimes R'$. Since $\sum_i P_i=\mathbbm{1}$,
we can replace $U$ with $U=\mathbbm{1} + \sum_{k\ne 0} (e^{i\delta_k}-1)P_k$. Inserting this into the previous equation we obtain
 \begin{equation}
 R\otimes R' \Big(\mathbbm{1} + \sum_{k\ne k_0} (e^{i\delta_k}-1)P_k\Big) R\otimes R'\;=\;
 R\otimes R'\,,
\end{equation}
 which further gives
\begin{equation}
 R\otimes R' \Big(\sum_{k\ne 0} (e^{i\delta_k}-1)P_k\Big) R\otimes R'\;=\;0\,.
\end{equation}
Suppose now that for each $k\ne 0$ we have $P_k=\sum_m \proj{\gamma_k^m}$.
Then, repeating the reasoning from Sect.~3, we arrive at
\begin{equation}
\forall\; k,m\; \forall\; i,j \quad \bra{i}U_1 C_k^m U_2^T \ket{j}\;=\;0\,,
\end{equation}
where $C_k^m$ are the Schmidt matrices of $\gamma_k^m$.
This means that the space of matrices $C_k^m$ must be $(d_1-M,d_2-N)$-decomposable.

Consider now a general random unitary channel given by 
(\ref{rotowany}) with $\varrho=\varrho_1\otimes\varrho_2$. The
equivalence between DFSs and UCCs still holds (notice that in
general a DFS for a channel (\ref{rotowany}) is a UCC for a
channel (\ref{ruc})). Now we need to solve
\beqn
R\otimes R'\dager{U_i}U_j R\otimes R'\;=\;\lambda_{i}^*\lambda_j R\otimes R'\,,
\eeqn
$i,j =0,1,\ldots$, with $\lambda_j=\eksp^{\uroj \varphi_j}$ and $U_0=\mathbbm{1}$. 
A solution may exist only when the unitaries $U_m$ ($m\ne 0$) are, up to an irrelevant global phase, of the form
\beqn
U_m\;=\;P_m+\sum_{k} e^{i\alpha^{m}_k}Q_k^m
\eeqn
with $P_i$ such that $P_i\calH\subseteq P_j\calH$ for $j>i$ and $r (P_1)\ge MN$ (if unitaries have the proper form we can always arrange them in that order). We then choose $P_1$ and check this subspace for decomposability.

Let us now discuss applicability of the previously used  technique to channels with general, i.e., not proportional to unitaries, Kraus operators $\{ A_i \}$.
  What we can do then is to find the largest standard, \tzn, nonproduct, DFC (which may not be easy itself) and use the results of this paper to check whether 
  it has a product subspace of required dimensionality. However, we have to remember that in the generic case the equivalence between DFS and UCC does not hold, thus for $k$-UC channels this procedure would only constitute the first step as we would still need to verify whether recovery can be performed locally.

The technique we introduced can also be useful for the case of a larger number of senders.
  Assume for simplicity that we have four of them denoted $A,B,C,D$. We seek for a product code $R_{AB}\otimes R_{CD}$ along the $AB|CD$ cut. If $R_{AB}$ and $R_{CD}$ are product themselves we are done. If there are no such $AB$, $CD$ projectors --- we conclude that there is no product code. Verification of which of the cases is true can be done by checking $A|BCD$, $ABC|D$ cuts.

 Finally, notice that each time when it is enough to perform product recovery for the product codes we construct, the same codes will be good for communication through the same channel but considered in a broadcast setup. These codes constitute a subset of all UCCs in these cases.

\section{Conclusions}

Summarizing, we have considered
construction of decoherence-free codes for random unitary channels with many inputs. We have started with
a hermitian noise model for a bi-unitary  two-access channel and have shown that the problem in this case was directly related to the characterization of decomposability properties in spaces of matrices of bounded
rank, an area of linear algebra, which is quite well established and understood. To the best of our knowledge this is the first time when analyses of spaces of matrices of bounded rank enter the field of quantum error correction. The result has been further extended to the case of arbitrary random unitary channels. We have also discussed applicability of the technique to channels with arbitrary Kraus operators and channels with an arbitrary number of inputs.

We encourage the reader to consult \cite{my} for methods of constructing higher entropy codes.

\section*{Acknowledgments}

Discussions with P. Horodecki and K. \.Zyczkowski are gratefully acknowledged. The author was supported by Gda\'nsk
 University of Technology through the grant ``Dotacja na kszta\l cenie m\l odych kadr w roku 2011".

\end{document}

%% file: komendy_osid.tex
\newcommand{\beq}[0]{\begin{equation}}
\newcommand{\eeq}[0]{\end{equation}}
\newcommand{\bw}[0]{\begin{widetext}}
\newcommand{\ew}[0]{\end{widetext}}
\newcommand{\bc}[0]{\begin{center}}
\newcommand{\ec}[0]{\end{center}}
\newcommand{\bwn}[0]{\begin{widetext}\begin{eqnarray}}
\newcommand{\ewn}[0]{\end{eqnarray}\end{widetext}}
\newcommand{\beqn}[0]{\begin{eqnarray}}
\newcommand{\eeqn}[0]{\end{eqnarray}}

\newcommand{\np}[0]{e.g.}
\newcommand{\tzn}[0]{i.e.}

\newcommand{\uroj}[0]{i}
\newcommand{\eksp}[0]{e}
\newcommand{\proj}[1]{|#1\rangle \langle #1|}
\newcommand{\ket}[1]{|#1\rangle}
\newcommand{\bra}[1]{\langle #1 |}
\newcommand{\dager}[1]{{#1}^{\dagger}}

\newcommand{\outerp}[2]{\ket{#1}\! \bra{#2}}
\newcommand{\inner}[2]{ \langle #1 | #2 \rangle}

\newcommand{\spann}[0]{\mathrm{span}}

\def\calC{{\cal C}}
\def\calD{{\cal D}}

\def\calH{{\cal H}}

\def\calL{{\cal L}}

\def\calP{{\cal P}}
\def\calQ{{\cal Q}}

\def\frakA{{\frak A}}
\def\frakB{{\frak B}}
\def\frakC{{\frak C}}
\def\frakD{{\frak D}}

\def\frakH{{\frak H}}